# Is the Born Rule Anthropically Determined?

M. B. Weissman

Department of Physics

University of Illinois at Urbana-Champaign

1110 West Green Street, Urbana, IL 61801-3080

Abstract: I argue that various derivations of the Born probability rule within strictly unitary quantum mechanics implicitly use anthropic constraints on the properties of the probabilities, but without proposing an ensemble from which those constraints limit a selection. An argument by Mallah obtains the Born probabilities explicitly for standard unitary dynamics but via postulating a hypothetical extra noise-like component of the quantum state. I argue that the implicit anthropic constraints required for well-behaved probability rules would select for a state of the sort Mallah proposes.

**Introduction**

The central problem in the interpretation of quantum mechanics is that the unitary time evolution unambiguously predicts outcomes which are supersets of the observed outcomes.[1] Observed outcomes have well-defined values of macroscopic parameters, while formal unitary outcomes do not. The relative frequencies of occurrence of observing any particular sets of macro outcomes are well known to follow the Born rule, i.e. to be proportional to the measure of the part of the formal outcome which has those properties. In traditional interpretations of quantum mechanics, a collapse process is more or less explicitly introduced whereby the quantum state shrinks to one or another of its macroscopically definite components, with the choice of which one survives following the Born probability rule. The Many Worlds Interpretation (MWI) dispenses with this unobserved non-unitary non-local collapse process.[1] Although the resulting picture is mathematically more coherent, it lacks precisely the ill-defined stage at which the Born rule can be introduced by fiat. Collapse-free accounts ought then to be able to predict the observed occurrence frequencies by some argument based directly on the formalism. Here I shall argue that it might do so only conditionally, with the conditions determined anthropically.

**Implicitly Anthropic Arguments in Many Worlds**

The MWI is typically introduced via simplified toy decoherence events, in which some initially coherent quantum state decoheres into a discrete list of macro-distinct components, each entangled with a different version of the environment. The obvious way to infer frequencies of occurrence in such a picture is simply to count "worlds", i.e.

distinct decoherent outcomes. That procedure is well-known to give nothing like the Born rule. (see e.g. [2]) Various arguments have been made to the effect that nonetheless the Born rule *must* be followed. (see, e.g. [3]) In general, those arguments take the form of assuming that the occurrence frequencies are describable by factorizable probabilities assigned to events, so, for example, the probability that Schrödinger's cat lives on Tuesday will not depend on whether it acquires quantum fleas on Wednesday. The probability then becomes a conserved flowing quantity. For non-trivial quantum spaces the quantum measure is the only such quantity. (see, e.g. [4] [5])

The problem with such an argument is that it assumes that probabilities must be well-behaved, in that the probability of Tuesday's outcomes does not change *after* Tuesday. In some versions of the argument, intersubjective agreement on a universal probability is also assumed (e.g. [5]), even though the need for a probabilistic description of decoherent components arises entirely from the division of the quantum events into an observer+observed part and an external environment part, so that the observer-independence of the probabilities is not an obvious given. Since in outcome-counting, Wednesday's fleas do alter the relative counts of live and dead cats after Tuesday's experiment[2], it's clear that the simplest outcome-counting interpretation cannot fit our observations.

Should we then simply say that outcome-counting has nothing to do with occurrence frequencies? By way of analogy, if in classical physics one were to grant the common-sense assumption that blackbody radiation is finite, one could derive the $T^4$ law. The problem would be that such an assumption makes no sense in the given framework, so that following this comforting procedure would be a way to avoid discovering quantum mechanics. One should examine whether there is any way to make sense of the observed outcomes in the framework of the fundamental theory, rather than simply assume that the implications of the framework must be reasonable.

Wallace[3] has made perhaps the most pertinent criticism of outcome-counting algorithms. He points out that under any realistic circumstances observers such as ourselves will never have even a single "world" if by world we mean a decoherent branch of the quantum state, i.e. a pure diagonal term in a density matrix, traced over the environment, which remains stable in a given basis for long enough to represent a state of mind. There's just far too much decoherence going on all the time. We're qualitatively dissimilar to the toy case used to initially introduce the MWI. Therefore we must construct some argument concerning the probabilities without hoping to count discrete branches. Wallace argues that simply adopting a measure-based probability for operator values is then justified. The inapplicability of simple count-based arguments does not, however, logically imply the applicability of measure-based arguments.[6]

To claim that the Born rule *must* hold in order for there to be objective factorizable probabilities does not seem adequate, since we have a nominally complete fundamental form of quantum theory, which ought to fully specify its own implications. One could argue that anthropic constraints require that any probabilities have the well-behaved properties of those we observe, in order to be accessible to anything like a mind, but

4                                          3/27/14

anthropic arguments cannot be justified unless there is some ensemble on which the anthropic selection occurs. Thus the anthropic argument that the probabilities must be reasonable would seem to be quite irrelevant to the Born rule, since the usual sorts of ensembles considered all share the Born rule. I shall argue that this impression is not necessarily correct.

**Mallah's Unitary Account of the Born Rule**

There have been several approaches to deriving the intrinsic non-linearity of the Born rule from the purely linear dynamics of the quantum formalism. The first, by Hanson[7], considered non-linear criteria for the experiencability of a "world" in a background of other components of a quantum state. Hanson's suggestion seemed unworkable in detail since there was no reason for the different decoherence processes in different macro-branches to match well enough to give a fixed ratio of sub-branch counts to net quantum measure, even if we accepted the unrealistic idea of discrete sub-branches.

Mallah [8] has proposed a different framework for generating a frequency ratio from measure. The frequencies to be counted, as in a variety of "many minds" versions of the Many Worlds approach, are the frequencies of independent thoughts. Mallah takes a fully functionalist approach, treating a thought as a robust computation performed by some subset of a quantum system. In any such computation the "signal" must exceed any irrelevant "noise". Mallah argues that if the quantum state consists of two components, one signal-like and the other noise-like, the Born rule can emerge directly from this consideration without introducing any non-unitary dynamics.

Mallah's argument goes roughly as follows. Say that the noise has a white spectrum in any standard coordinate representation. In order for the signal to beat the noise, one must average over enough of the coordinate space. Standard statistics then requires averaging over a volume inversely proportional to the squared magnitude of the signal. Thus the ratios of numbers of robust independent computations corresponding to different outcomes would obey the Born rule, regardless of the somewhat arbitrary resolution criteria used to describe which computations are independent.

Perhaps the most obvious difficulty with Mallah's idea is that there is no independent reason to postulate that the existence of the white-noise background component. On the other hand, the initially low-entropy component of which we usually keep track itself has uncertain origins. Modern cosmology has not entirely obviated Boltzmann's early speculation[9] that the explanation for the statistical peculiarity of our state might ultimately be anthropic.(See discussion in [10].)

**Combining Mallah with Anthropic Constraints**

So let us now put together the ingredients. The usual arguments that the Born probabilities must hold are implicitly though not explicitly anthropic. They lack any explanatory mechanism with parameters on which anthropic post-selection might occur. Mallah's explanation has a mechanism with an adjustable collection of properties, those of the "noise" component of the starting state, but lacks any account of why just the right sort of component is found to give Born probabilities. I am proposing the obvious combination of these approaches.

Only Born probabilities have the sort of properties that a rational actor might be able to track, in particular the factorization of probabilities for sequential events. Our only good account (other than assertions by fiat) for the emergence of Born probabilities in unitary quantum mechanics requires that Mallah's hypothetical background noise be part of the state. Therefore standard anthropic reasoning would require that of the ensemble of all states (all presumably realized in some abstract space) only the Mallah-style states could be experienced by anyone who could begin to discuss such questions. Therefore that's the type of state in which we conduct this discussion.

Does this proposal accomplish anything beyond tying together some loose ends in the Many Worlds interpretation? At an even more speculative level, it may connect with another prominent loose end in cosmology. A wide variety of cosmological models have trouble accounting for why our universe is flat and homogeneous on the widest observable scales, far beyond known anthropic requirements. (See [11] for an informal discussion.) If the Born rule itself is anthropically required and derived from actual physical interaction with statistically homogeneous quantum fields propagating at up to the speed of light, anthropic constraints on large-scale homogeneity might extend to horizons.

The first obvious question for any such proposal is whether it has observational implications beyond retrodicting the already known properties of the world. I cannot yet specify any such new predictions. Nevertheless, any scenario in which the Born probabilities emerge from contingent properties of the quantum state allows in principle for the possibility that they will not be exact and universal.

**Acknowledgements**: I thank Jacques Mallah for stimulating conversations, but do not wish to imply his agreement with my proposal.